\documentclass[twocolumn,prl,showpacs,preprintnumbers,amsmath,amssymb,a4paper,fleqn]{revtex4-1}

\usepackage{dcolumn}
\usepackage{epsfig}
\usepackage{amssymb,amsmath,bm,graphicx,multirow,mathtools}
\usepackage{color}

\def\be{\begin{equation}}
\def\ee{\end{equation}}
\def\bea{\begin{eqnarray}}
\def\eea{\end{eqnarray}}
\def\beeg{\begin{align}}
\def\eeg{\end{align}}

\def\m{\mu}
\def\n{\nu}

\def\p{\partial}
\def\a{\alpha}
\def\b{\beta}
\def\t{\theta}

\def\11{1\hspace{-0.6em}1}
\def\22{1\hspace{-0.48em}1}
\def\00{0\hspace{-0.45em}0}

\def\tt{\theta\hspace{-0.45em}\theta}
\def\pl{\p\hspace{-0.5em}^{^{^\leftarrow}}}
\def\pr{\p\hspace{-0.5em}^{^{^\rightarrow}}}

\def\xx{\mathbf{x}}
\def\yy{\mathbf{y}}

\def\yyh{\hat{\yy}}

\def\i{\imath}

\def\g{\gamma}
\def\tt{\theta\hspace{-0.43 em}\theta}
\begin{document}

\title{The effect of nucleus size on the electron energy levels: via Seiberg-Witten map}
\author{Abolfazl\ Jafari\footnote{jafari-ab@sci.sku.ac.ir}}
\affiliation{Department of Physics, Faculty of Science,
Shahrekord University, P. O. Box 115, Shahrekord, Iran}
\date{\today }

\begin{abstract}

\textbf{Abstract:} 
We compare the effects of the entering non - commutative geometry in physics, which have studied by Bob's shift method and Seiberg-Witten map.
Due to the corrections of the electron energy levels, we demonstrate that two approaches are not equivalent. 
We show that the electric and magnetic dipole moments, as well as one loop vertex correction, change due to the mapping selection.
Furthermore, we provide a way to communicate the results of the two methods.

\end{abstract}


\maketitle


\section{Introduction}

Seiberg-Witten map is one of the methods of introducing the non-commutative geometry in physics.
The principles of this map have explained in many texts \cite{witten, douglas, gitman}.
The most instances of the studies of physicists in the field of non-commutative geometry are the researches on the effects of importing non-commutative coordinates, which are mainly limited to the first order of the non-commutative parameters.
One of the conventional methods in rewriting the physics on the non-commutative geometry is Moyal-Weyl mapping \cite{nekrasov, szabo1, szabo2, jab1, jab2}.
However, Moyal-Weyl mapping is also used up to the first order of non-commutativity.
The non-commutative geometry will showing-off if we accept that the following commutativity
\begin{eqnarray} \label{canonical commutation relation}
&&[\hat{y}^\m,\hat{y}^\n]_\star=\i\t^{\m\n},
\end{eqnarray}
holds for the coordinates.
Here, $\t$ as a real constant and anti-symmetric tensor has a square length dimension.
%Further, $[\breve{A},\breve{B}]_\star=\breve{A}\star\breve{B}-\breve{B}\star\breve{A}$.
The sign of product $'\star '$, is a method for the formulation of the physics in the non-commutative geometry.
A simple rule explains how we should apply it.
Indeed, replacing the ordinary products between quantities with a $\star$-product \cite{connes, wess, szabo, bertolami, jaf4},
\begin{align}\label{mw}
\breve{f}(\yyh)\breve{g}(\yyh)=\breve{f}(\yy)\star\breve{g}(\yy)=\breve{f}(\yy){\rm e}^{\frac{\i}{2}\pl_{\ \ \m}\t^{\m\n}\pr_{\ \ \n}}
\breve{g}(\yy),
\end{align}
where $\p_\a=\frac{\p}{\p y^\a}$.
In most cases, because of the causality time coordinate does not contribute to the non-commutativity.
According to the literature, the mean value of the non-commutativity parameter is in order of $\sqrt{\t}\sim 10^{-15}\mathrm{m}$ \cite{saha, jab3}.
Based on Eq.(\ref{canonical commutation relation}), at the first order of $\t$, the variation of electromagnetic fields can be explained by the SW map:
\begin{align}\label{006}
\breve{A}_\m=A_\m-\frac{1}{2e}\t^{\a\b}A_\a(\p_\b A_\m+f_{\b\m}),
\end{align}
where 
%$g=\frac{e}{c\hbar}$ and 
$f_{\m\n}=\p_\m A_\n-\p_\n A_\m$.
Here, we use the hat symbol to specify the functions containing the non-commutative coordinates.
By substituting of Eq.(\ref{canonical commutation relation}), the changes of the field strength tensor $\breve{F}^{\m\n}=\p^\m\breve{A}^\n-\p^\n\breve{A}^\m-\frac{\i}{e}[\breve{A}^\m,\breve{A}^\n]_\star$, read
\begin{align}\label{0011}
F^{\m\n}=f^{\m\n}-\frac{1}{e}\t^{\a\b}\{A_\a\p_\b f^{\m\n}-f_\a^{\ \ \m} f_\b^{\ \ \n}\},
\end{align}
where the equation of motion for the strength tensor as following
\begin{align}\label{007}
\breve{\mathfrak{D}}_\m\star\breve{F}^{\m\n}(\yyh)=\frac{4\pi}{c}\breve{j}^\n(\yyh).
\end{align}
Moreover, the covariant derivative defined by
\begin{align}
\breve{\mathfrak{D}}_\m\star=\p_\m-\frac{\i}{e}[\breve{A}_\m,\star].
\end{align}
The four-vector of the current also is changed in the SW map.
The current term transforms by $\breve{\acute{j}}^\m=u_\star(\breve{\Lambda})\star\breve{j}^\m\star u^{-1}_\star(\Lambda)$ in which $\Lambda$ should be changed:
\begin{align}
\breve{\Lambda}=\Lambda-\frac{1}{2e}\t^{\a\b}A_\a\p_\b \Lambda.
\end{align}
So, the variation rule for the four-vector of the current is written as
\begin{align}\label{000}
\breve{j}^\m=j^\m-\frac{1}{e}\t^{\a\b}A_\a\p_\b j^\m.
\end{align}

In this work, we admit the non-commutative geometry and restrict ourselves to the first order of $\t$.
We know that the SW map is only valid when we have $\t^{\m\n}\t_{\m\n}>0$ and this condition realized in this paper. 
We also calculate the corrections of the Maxwell equations and make deformed electrodynamics based on the SW map. 
As a main of this paper is proving that the results of the two approaches are different.
Moreover, we want to show that the nucleus size effects on the energy levels of the electron layers, emission lines and absorption of atoms.

\section{Basic notations}

In a semi classical interpretation, the center of atoms has a charge distribution function that lies within the nucleus position which in this region the electron cannot be seen.
The shape of a nucleus can be assumed to be a sphere with radius $a$.
As we mentioned in the previous section, the radius of the nuclear is at least larger than $\sqrt{\t}$.
So, $a\gtrsim10^{-15}\mathrm{m}$ which has a reference to the proton diameter.
We denote the region $r<a$ and its related quantities with the index of $_{_<}$ and by the same style also for the region $r>a$ with $_{_>}$.
We can split the electromagnetic fields into the two separate parts, $A^\m=A^{(0)\m}+A^{(1)\m}+0(\t)^2$,
which the second term is small.
Assuming spherical nucleus and symmetric distribution of electric charge we can write
\begin{align}
j^{(0)0}=\rho(r)=\frac{3ze}{4\pi a^3},\ r<a,\ \ j^{(0)0}=0,\ r>a,
\end{align}
which mentions the components potential function as follows
\begin{align}
A_<^{(0)0}=-\kappa_{_<} r^2+3\kappa_{_<}a^2,\ \ \ \ A_>^{(0)0}=\frac{\kappa_{_>}}{r},
\cr
A^{(1)0}_{<(>)}=0.
\end{align}
where $\kappa_{_<}=\frac{ze}{2 a^3}$ and $\kappa_{_>}=ze$.
According to Eq.(\ref{0011}), Eq.(\ref{007}) yields the following equation
\begin{align}
\p_\n f^{\n\m}-\frac{1}{e}\t^{\a\b}\p_\n(A_\a\p_\b f^{\n\m}-f_\a^{\ \ \n} f_\b^{\ \ \m})
\cr
+\frac{1}{e}\t^{\a\b}\p_\a A_\n\p_\b f^{\n\m}
=\frac{4\pi}{c}j^{\m}.
\end{align}
Based on the initial selected conditions, the gauge $\p_i A^{(0)i}=0$ and the equation of $j^{(1)\m}=0$, is realized.
In this way, vector potentials $A_{_{<(>)}}$ obey the following:
%we obtain the following independent equations
\bea\label{111}
&&\nabla^2 A^{(1)k}_{_<}(\xx)=4\frac{1}{e}\kappa^2_<\t^{ik}x_i,
\nonumber\\&& 
\nabla^2 A^{(1)k}_{_>}(\xx)=\frac{1}{e}\kappa^2_>\t^{ik}\frac{x_i}{r^6}.
\eea
Solving Eqs.(\ref{111}) must be due to the boundary conditions:
$A^{(0)\m}_{_<}(\xx)=A^{(0)\m}_{_>}(\xx)|_{r=a}$ and $\p_r A^{(0)\m}_{_<}(\xx)=\p_r A^{(0)\m}_{_>}(\xx)|_{r=a}$.
Therefore, the solutions of Eqs.(\ref{111}) are
\bea
&&A^{(1)k}_{_<}(\xx)=\frac{1}{e}\frac{4}{10}\kappa^2_<\t^{ik}x_ir^2+\eta_{_<}(r,\t,\phi),
\nonumber\\&&
A^{(1)k}_{_>}(\xx)=\frac{1}{e}\frac{1}{4}\kappa^2_>\t^{ik}\frac{x_i}{r^4}+\eta_{_>}(r,\t,\phi),
\eea
where $\nabla^2\eta_{_{<(>)}}(r,\t,\phi)=0$ and we have infinite solutions.
We choose the following solutions
\bea\label{outsol}
&&A^{(1)k}_{_<}(\xx)=\frac{1}{e}(\frac{ze}{2a^2})^2(\frac{2r^2}{5a^2}-1)\t^{ik}x_i,
\nonumber\\&&
A^{(1)k}_{_>}(\xx)=-\frac{1}{e}\frac{1}{4}(\frac{ze}{r^2})^2(\frac{8r}{5a}-1)\t^{ik}x_i,
\eea
which satisfy the requirements of the Lorentz gauge condition.

\section{Presentation of theory}

In the following, our calculations are only in the region $r>a$.
Thus, we can drop the separator indicator from this area.
In addition, all the quantities that come after will be belonging to the area $r>a$.
For the electrons from the region $r>a$, the Hamiltonian $\hat{H}$ is given by
\begin{align}
\hat{H}=\frac{\hat{\mathbf{\Pi}}^2}{2m}+\hat{V}(r).
\end{align}
where, $\Pi^k=p^k+\frac{e}{c}A^{(1)k}$.
In the first order of $\t$, the operator of momentum becomes, $\hat{\mathbf{\Pi}}^2=\hat{\mathbf{p}}^2+\frac{2e}{c}A^{(1)k}\hat{p}_k-\i \hbar\frac{e^2}{c^2}\p_k A^{(1)k}+0(\t)^2$.
Since $\nabla\cdot A^{(1)}$ is zero, then
the Hamiltonian is summarized as
\begin{align}
\hat{H}=\frac{\hat{\mathbf{p}}^2}{2m}+V_{eff}(r;\t,a),
\end{align}
where
\begin{align}\label{0012}
V_{\mathrm{eff}}(r;\t,a)=-\frac{ze^2}{r}-\frac{1}{4mc}z^2e^2f(r;a)\tt\cdot\mathfrak{L},
\end{align}
where $\tt=(\t^{23},\t^{13},\t^{12})$ is a new vector and $f(r;a)=\frac{8}{5ar^3}-\frac{1}{r^4}$.
It is evident that the effective potential contains the nucleus size and is different from the potential offered in Ref.{\citep{jab3}}.
In the Ref.{\cite{jab3}}, the authors have found an effective potential as 
\begin{align}\label{0013}
V_{\mathrm{eff}}^{\mathrm{Bob's\ shift}}(r;\t)=-\frac{ze^2}{r}-\frac{ze^2}{4\hbar r^3}\tt\cdot\mathfrak{L},
\end{align}
by exploiting the Bob's shift method.
From the perturbation theory, the correction of the energy levels is
\begin{align}\label{0014}
\Delta E_{NC}=<n\acute{l}j\acute{j}_z|\hat{H}_{p}|nljj_z>,
\end{align}
where the perturbed Hamiltonian is given by the second parts of Eq.(\ref{0012}) or Eq.(\ref{0013}).
In the following, we set; $$\tt=(0,0,\t)$$ because of the simplicity in calculations.
For the chosen case of the non-commutativity, 
by substituting the second part of Eq.(\ref{0012}) in Eq.(\ref{0014}), Eq.(\ref{0014}) becomes
\begin{align}\label{0015}
\Delta E^{SW}_{NC}=
-\frac{z^2e^2}{4mc}f_{nl}(r;a)j_z\hbar(1\mp\frac{1}{2l+1})\delta_{\acute{l}l}\delta_{\acute{j}_zj_z},
\end{align}
in which 
$$f_{nl}(r;a)=<nl|f(r;a)|nl>=\frac{8}{5a}<r^{-3}>-<r^{-4}>$$ and the condition $j=l\pm\frac{1}{2}$ is considered.
By holding the term of $<r^{-3}>$ and replacing the value of
%\begin{align}
%\begin{split}
%%\label{eq:2}
%{}&
%<r^{-3}>=\frac{8z^3}{r_0^3}\frac{(2l-1)!}{(2l+2)!},
%\\ &
%<r^{-4}>=\frac{2^4z^4(2l-2)!}{r_0^4n^5(2l+3)!}(3n^2-l(l+1)).
%\end{split}
%\end{align}
\begin{align}
<r^{-3}>=\frac{8z^3}{r_0^3}\frac{(2l-1)!}{(2l+2)!},
\end{align}
the Eq.(\ref{0015}) gives 
\begin{align}\label{005}
\Delta E^{SW}_{NC}=
-\frac{8ze^2\hbar}{5amc}f_{nl}(r)j_z(1\pm\frac{1}{2l+1})\delta_{\acute{l}l}\delta_{\acute{j}_zj_z},
\end{align}
in which $r_0$ indicates the radius of Bohr.
With, $e^2r_0^{-3}=z^3mc^2\a^4\lambda_c^{-2}$, we have
\begin{align}
\Delta E^{SW}_{NC}=
\cr
-\frac{8z^4\hbar c\a^4\lambda_c^{-2}\t}{20a}
g_{nl}(r)j_z(1\mp\frac{1}{2l+1})
\delta_{\acute{l}l}\delta_{\acute{j}_zj_z}.
\end{align}
where $g_{nl}(r)=r^3_0<r^{-3}>$.
At the same level, the correction function from Ref.{\cite{jab3}}, based on Eq.(\ref{0013}), is 
\begin{align}
\Delta E^{MW}_{NC}=
\cr
-\frac{z^4mc^2\a^4\lambda_c^{-2}\t}{4}g_{nl}(r)j_z(1\pm\frac{1}{2l+1})\delta_{\acute{l}l}\delta_{\acute{j}_zj_z},
\end{align}
We calculate the ratio of the corrections
\begin{align}
\frac{\Delta E^{SW}}{\Delta E^{MW}}=\frac{8\hbar}{5amc}\propto10^{3}.
\end{align}
This ratio shows that the magnitude of correction due to SW mapping is $10^{3}$ times the one caused by Bob's shift.
Therefore, the correction of the electron energy levels in a hydrogen-like atom calculated by SW map and Bob's shift methods, are not equivalent to one another.
By comparing the potential of Eq.(\ref{0012}) and Eq.(\ref{0013}) and regardless of the term $<r^{-4}>$ in Eq.(\ref {0012}), it can be shown that by replacing $\t$ with ${8\hbar\t/5amc}$ in the results of Ref.\cite{jab3}, the electric and magnetic dipole moments of the electron are changed as
\bea
&&<\vec{\m}>=-\frac{e}{2mc}(g\vec{s}+\frac{8\a\g_E}{15amc\pi}\frac{\hbar^2}{\lambda^2}\t),
\nonumber\\&&
<\vec{P}>=-\frac{e}{4\hbar}(\frac{8}{5amc}\mathbf{\t}\times\mathbf{P})(1+\frac{3\a\g_E}{m}).
\eea
Furthermore, the change of the vertex correction in the one loop order is 
\begin{align}
V^{one\ loop}_{NC\ vertex}=-\frac{8ze^2}{20amc\pi}\a\g_E(3-\frac{2}{3})\frac{\mathfrak{L}\cdot\mathbf{\t}}{r^3}.
\end{align}
The changes are quite evident from the Bob's shift method.

\section{Conclusion}

At the level of quantum mechanics, we found a way to determine the contribution of the size of atomic nuclei in the electron energy levels.
We extracted the Hamiltonian of system at the classical mechanics level by exploiting Seiberg-Witten map based on the $\t$ deformed electrodynamics.
We showed that the Hamiltonian changes from the official version by entering the size of the nucleus in the electrodynamics equations.
It was determined the dependence of the energy corrections on the orbital angular momentum of the electron and the size of the atomic nucleus.
We also proved that the energy corrections from the SW maps and Bob's shift method are different, and their ratio is in the order of $10^3$.
Based on this paper, the electric and magnetic dipole moments of the electron as well as vertex function in the order of one loop, have many variations that come from the amount of their values given in Ref.\cite{jab3}, with $\t^{MW}\rightarrow\t^{SW}=\frac{5amc}{8\hbar}\t^{MW}$ substitution.

\section{Acknowledgments}
The author thanks Shahrekord University for supporting this work with a research grant.
\newline
%%%%%%%%%%%%%%%%%%%%%%%%%%%%%%%%%%%%%%%%%%%%%%%%%%

\end{document}